\documentclass[twocolumn,british,english]{revtex4-1}
\usepackage[T1]{fontenc}
\usepackage[latin9]{inputenc}
\usepackage{amsmath}
\usepackage{amssymb}
\usepackage{graphicx}

\makeatletter
\usepackage{babel}

\usepackage{babel}

\makeatother

\usepackage{babel}
\begin{document}

\title{Computing equilibrium states of cholesteric liquid crystals in elliptical
channels with deflation algorithms}

\author{David B. Emerson}

\affiliation{Department of Mathematics, Tufts University, 503 Boston Ave., Medford,
Massachusetts 02155, USA}

\author{Patrick E. Farrell}

\affiliation{Mathematical Institute, University of Oxford, Woodstock Road, Oxford
OX2 6GG, UK}

\author{James H. Adler}

\affiliation{Department of Mathematics, Tufts University, 503 Boston Ave., Medford,
Massachusetts 02155, USA}

\author{Scott P. MacLachlan}

\affiliation{Department of Mathematics and Statistics, Memorial University of
Newfoundland, St. John's A1C 5S7, Newfoundland, Canada}

\author{Timothy J. Atherton}
\email{timothy.atherton@tufts.edu}

\affiliation{Department of Physics and Astronomy, Tufts University, 574 Boston
Ave., Medford, Massachusetts 02155, USA}
\begin{abstract}
We study the problem of a cholesteric liquid crystal confined to an
elliptical channel. The system is geometrically frustrated because
the cholesteric prefers to adopt a uniform rate of twist deformation,
but the elliptical domain precludes this. The frustration is resolved
by deformation of the layers or introduction of defects, leading to
a particularly rich family of equilibrium configurations. To identify
the solution set, we adapt and apply a new family of algorithms, known
as deflation methods, that iteratively modify the free energy extremisation
problem by removing previously known solutions. A second algorithm,
deflated continuation, is used to track solution branches as a function
of the aspect ratio of the ellipse and preferred pitch of the cholesteric. 
\end{abstract}
\maketitle

\section{Introduction}

Cholesteric liquid crystals are complex fluids that exhibit long-range
orientational order, elasticity, local alignment at surfaces, optical
activity and response to external stimuli~\cite{bahr2001chirality}.
They are composed of chiral molecules that, in the absence of boundaries,
adopt a helical structure with a preferred pitch, $q_{0}$, set by
the molecular structure and the ambient temperature. There has recently
been a great deal of interest in cholesterics in confined geometries
because of parallels with other condensed matter systems such as chiral
ferromagnets, Bose-Einstein condensates and the Quantum Hall effect.
All of these systems exhibit topological defects under confinement,
including skyrmions and torons, where the boundary conditions preclude
adoption of the energetically preferred uniformly twisted state. Hence,
they are \emph{geometrically frustrated}. It was recognised some time
ago that nematic liquid crystals also may potentially form skyrmions,
but these are only metastable due to the lack of preferred twist \cite{bogdanov2003skyrmions}.

Liquid crystals are particularly attractive to study these defect
structures, because they can be conveniently produced and imaged in
three dimensions with techniques such as confocal microscopy \cite{smalyukh2010three}.
Cholesterics may form skyrmion lattices in two dimensions \cite{fukuda2011quasi}.
In three dimensions, torons resemble particulate inclusions \cite{ackerman2014two,leonov2014theory}
and form chains or lattices \cite{ackerman2015self}. Other more
complicated defect structures called ``twistions'' occur in films
thinner than the cholesteric pitch \cite{ackerman2016reversal}. They
also provide an elegant experimental actualisation of the Hopf fibration,
a map from the 3-sphere onto the 2-sphere such that each distinct
point of the 2-sphere comes from a distinct circle of the 3-sphere
\cite{chen2013generating}. Strongly confined geometries such as micropatterned
surfaces \cite{cattaneo2016electric}, channels \cite{kim2015periodic,guo2016cholesteric}
and droplets \cite{orlova2015creation} can all be used to control
and order the location of skyrmions.

A key challenge in simulating these systems is that, due to the geometric
frustration, they possess a particularly rich family of stationary
solutions of the free energy. The ground state strongly depends on
the shape of the domain and material parameters, including the cholesteric
pitch. Typically, extremisation is performed from an initial guess
using a relaxation algorithm or by solving a set of nonlinear Euler-Lagrange
equations. In either case, having found a solution, the question remains:
are there others? It is also highly desirable to track the solution
set as a function of geometric and material parameters to assemble
a bifurcation diagram.

A common approach to identify distinct solutions, referred to in the
mathematics literature under the umbrella term of multistart methods
\cite{Marti1}, is to begin from a significant number of initial guesses.
This requires extensive knowledge of the problem to devise a suitable
set of initial guesses and can be inefficient as multiple guesses
often converge to the same configuration. Other well-established techniques
include numerical continuation \cite{Chao1,Allgower1}, which is particularly
effective in fully resolving connected bifurcation branches but can
neglect solutions if they are not homotopic with respect to the continuation
parameters \cite{Farrell4}, and approaches, such as Branin's method,
that rely on numerical integration of the Davidenko differential equation
corresponding to the original nonlinear problem \cite{Davidenko1,Branin1}.
Branin's method is capable of systematic computation of multiple solutions
but requires determinant calculations that become prohibitively expensive
for large-scale problems without special structure.

In this paper, we adapt a new technique known as the \emph{deflation}
method \cite{Farrell1,adler2017combining} to a model problem in this
class, the configuration of a cholesteric in an elliptical channel.
The method is generalisable, robust and computationally efficient
for large-scale applications. It has been successfully applied to
a diverse set of nonlinear problems including nonlinear partial differential
equations (PDEs), singularly perturbed problems, the analysis of Bose\textendash Einstein
condensates, and the computation of disconnected bifurcation diagrams
\cite{Farrell1,Farrell5,Farrell6,Farrell4}. This paper is structured
as follows: in Section \ref{sec:Model}, we briefly describe the problem;
in Section \ref{sec:Deflation}, we introduce the deflation technique
with an illustrative toy example. In Section \ref{sec:Cholesterics-with-Curved},
we present results for the cholesteric problem. Conclusions are drawn,
with prospects for further applications of the algorithm, in Section
\ref{sec:Conclusion}.

\section{Model\label{sec:Model}}

We consider a cholesteric liquid crystal in a channel with elliptical
cross section. Equilibrium structures are found by identifying critical
points of the Frank energy, 
\begin{align}
F & =\frac{1}{2}\int_{\Omega}dV\ K_{1}\left(\nabla\cdot\mathbf{n}\right)^{2}+K_{2}\left(\mathbf{n}\cdot\nabla\times\mathbf{n}+q_{0}\right)^{2}+\label{eq:Frank}\\
 & +K_{3}\left|\mathbf{n}\times\nabla\times\mathbf{n}\right|^{2}\nonumber 
\end{align}
where $K_{1}$, $K_{2}$ and $K_{3}$ are the splay, twist and bend
elastic constants; $\mathbf{n}$ is a headless unit vector, the \emph{director,}
that corresponds to the local symmetry axis of the molecular orientational
distribution and $q_{0}$ is the preferred pitch for the cholesteric.
Rigid anchoring (Dirichlet) boundary conditions are imposed on the
boundary $\partial\Omega$, where the director is required to point
perpendicular to the plane of the cross section. The energy is readily
non-dimensionalised by introducing a typical length $\lambda$ and
dividing through by a characteristic magnitude of the elastic constants
$\tilde{K}$; henceforth, we use dimensionless parameters.

As discussed in the introduction, this problem promotes the existence
of multiple local equilibria by construction. To see why, first consider
the cholesteric in the absence of boundaries. As is well-known, the
minimiser of (\ref{eq:Frank}) is a unique uniformly twisted state.
Level sets of $\mathbf{n}$ form families of equally spaced planes
often referred to as cholesteric ``layers.'' We note a valuable
discussion of the limitations of this view is found in \cite{beller2014geometry}.
Variation away from this preferred structure, which is equivalent
to compressing or bending the layers, implies an elastic cost. Considering
a cholesteric in a disk, the minimisers of (\ref{eq:Frank}) are solutions
where $\mathbf{n}$ rotates about the radial axis and lies everywhere
perpendicular to it. The number of rotations is determined by the
cholesteric pitch, $q_{0}$, which promotes a constant rotation rate,
and level sets of constant orientation therefore form equally spaced
concentric circles.

For an elliptical domain, however, it is not possible to fill the
ellipse with equally spaced layers, and so defects or a variable layer
spacing must be introduced. The cholesteric order, which prefers a
uniformly twisted state, and the shape of the domain are in competition,
so the system is said to be \emph{geometrically frustrated}. The frustration
is resolved by adopting a compromise state, incorporating some combination
of layer deformation or defects; in common with other frustrated systems
there is typically more than one way to do this, leading to the possibility
of more than one minimiser.

Further, we explore solutions where $K_{2}>K_{1},K_{3}$, which might
occur in exotic liquid crystal systems~\cite{le2015large}. This
choice of parameters leads to a material that is \emph{doubly frustrated}
because it is required to twist by the cholesteric term but
the twist is relatively expensive compared to other deformation modes.
As a result, the cost of modulating the cholesteric layers is reduced.
The interaction of geometric and internal frustration is expected
to lead to a particularly rich solution set, because they permit multiple
ways of relieving the frustration: one solution might accommodate an
incommensurate number of cholesteric periods by folding the layers;
another might introduce a defect. These parameters therefore yield
an extremisation problem that we anticipate \emph{a priori} to be
very challenging to explore by naive multistart methods.

\section{Deflation\label{sec:Deflation}}

\begin{figure}
\begin{centering}
\includegraphics{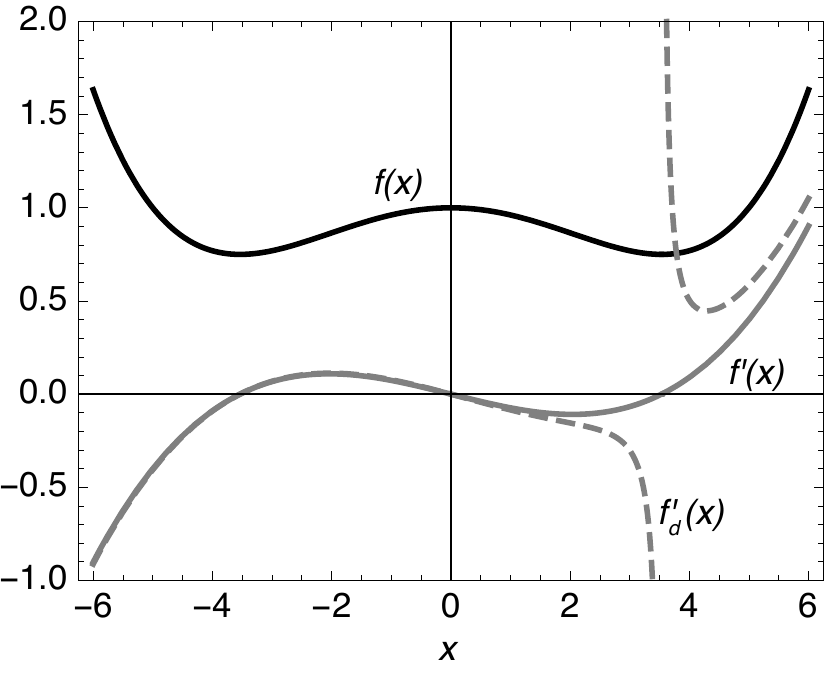} 
\par\end{centering}
\caption{\label{fig:Toy-1D-example}\textbf{Toy 1D example of deflation.} Critical
values of the function $f(x)$ (black), found by solving $f'(x)=0$
(gray). Having located a minimum at $x_{0}=\sqrt{5}/2$, a deflated
function $f'_{d}(x)$ (gray; dashed) is constructed; this now contains
a pole at $x_{0}$ but retains zeros in common with $f'(x)$. }
\end{figure}

In solving problems that possess multiple equilibrium solutions, such
as that posed in Section \ref{sec:Model}, a key challenge is to ensure
that the true ground state has been found from the set of energetically
low-lying solutions. The idea of the deflation algorithm is to successively
modify the nonlinear problem under consideration to eliminate known
solutions, enabling the discovery of additional distinct solutions.
Consider $\mathcal{A}(\mathbf{u})=\mathbf{0}$, a set of nonlinear
equations, that may admit multiple solutions. This system, for instance,
could represent a set of continuous or discretised nonlinear PDEs;
here we minimise the Frank energy in \eqref{eq:Frank}, subject to
the unit-length constraint on the director, and consider the resulting
system of nonlinear first-order optimality conditions. Using a known
solution, $\mathbf{v}$, to the system $\mathcal{A}(\mathbf{v})=\mathbf{0}$,
we define the deflation operator, 
\begin{align}
M_{p,\alpha}(\mathbf{u};\mathbf{v})=\left(\frac{1}{\left\Vert \mathbf{u}-\mathbf{v}\right\Vert _{U}^{p}}+\alpha\right)\mathbf{I},\label{DeflationOperator}
\end{align}
where $\alpha\geq0$ is a shift scalar, $p\in[1,\infty)$ is the deflation
exponent, and $\mathbf{I}$ is the identity operator. An appropriate
norm $\left\Vert \cdot\right\Vert _{U}$ must be chosen for the vector
space to which the solutions belong. The deflated system is then formed
by applying the deflation operator to the original nonlinear system
as, 
\begin{equation}
\mathcal{B}(\mathbf{u})=M_{p,\alpha}(\mathbf{u};\mathbf{v})\mathcal{A}(\mathbf{u})=\mathbf{0}.
\end{equation}
Iterative techniques, such as Newton's method, may then be applied
to solve the deflated system. While these iterations are guaranteed
to \emph{not} converge to the known solution $\mathbf{v}$ under mild
regularity conditions, the remainder of the solution space for the
original system is preserved by the deflation operator. Numerical
experiments have found the effectiveness of the deflation operator
to be relatively insensitive to the choice of deflation parameters.
However, for certain problems, performance improvements may be obtained
by varying $p$ and $\alpha$ \cite{Farrell1,adler2017combining}.
Typical values, used everywhere in this paper, are $\alpha=1$ and
$p=2$.

Having found two solutions $\mathbf{v}_{1},\mathbf{v}_{2}$, an expanded
deflation operator can be constructed by composition of single-solution
deflation operators, 
\begin{equation}
M_{p,\alpha}(\mathbf{u};\mathbf{v}_{1},\mathbf{v}_{2})=M_{p,\alpha}(\mathbf{u};\mathbf{v}_{1})M_{p,\alpha}(\mathbf{u};\mathbf{v}_{2}),
\end{equation}
and applied to the original nonlinear system. As the set of known
solutions $\{\mathbf{v}_{1},\mathbf{v}_{2},...,\mathbf{v}_{m}\}$
is expanded, the deflation operator is grown as the product of the
single deflation operators for each distinct solution in the set,
\begin{equation}
M_{p,\alpha}(\mathbf{u};\mathbf{v}_{1},\mathbf{v}_{2},...,\mathbf{v}_{m})=\prod_{i=1}^{m}M_{p,\alpha}(\mathbf{u};\mathbf{v}_{i}),\label{eq:deflationoperator}
\end{equation}
and its action remains multiplicative on the original system.

To provide a simple and tractable illustration of the deflation process,
we apply it to the problem of locating critical values of a one dimensional
objective function, 
\begin{equation}
f(x)=\frac{1}{5^{4}}x^{4}-\frac{1}{5^{2}}x^{2}+1,
\end{equation}
displayed in Fig. \ref{fig:Toy-1D-example} by solving the equation,
\begin{equation}
f'(x)=\frac{4}{5^{4}}x^{3}-\frac{2}{5^{2}}x=0.\label{eq:derivative}
\end{equation}
Starting from the initial guess $x=2.2$, Newton's method locates
the first solution $x_{0}=\sqrt{5}/2$. The deflation operator is
constructed following the definition in (\ref{DeflationOperator}),
as 
\begin{equation}
M_{p,\alpha}(x;x_{0})=\frac{1}{\left|x-x_{0}\right|^{p}}+\alpha,
\end{equation}
where $\left|\cdot\right|$ denotes the standard absolute value. Applying
this to \eqref{eq:derivative} yields the deflated optimality condition,
\begin{eqnarray}
0=f'_{d}(x) & = & M_{p,\alpha}(x;x_{0})f'(x)\nonumber \\
 & = & \left(\frac{1}{\left|x-x_{0}\right|^{p}}+\alpha\right)\left(\frac{4}{5^{4}}x^{3}-\frac{2}{5^{2}}x\right),\label{eq:deflated}
\end{eqnarray}
to be solved for $x$. The function $f'_{d}(x)$ is also plotted in
Fig. \ref{fig:Toy-1D-example} for deflation parameters $\alpha=1$
and $p=2$. Notice that $x_{0}$ is \emph{not} a solution to $f'_{d}(x)=0$,
while the remaining solutions to $f'(x)=0$ persist as solutions to
$f'_{d}(x)=0$. Use of the deflation operator precludes convergence
of certain iterative techniques, such as Newton's method, to $x_{0}$
while facilitating convergence to additional distinct solutions from
the same initial guess. With the deflation parameters chosen previously
and the same initial point, $x=2.2$, Newton's method converges to
the critical point $x_{1}=0.0$. Thus, two solutions are obtained
from the same initial guess. The process may then be repeated by constructing
a multi-deflation operator, incorporating both known roots, to enable
the discovery of the third distinct solution at $x_{2}=-\sqrt{5}/2$
to (\ref{eq:derivative}) and hence identify all extremal values of
$f(x)$.

While deflation is a useful device for finding distinct solutions,
the number of solutions discovered may still depend on the analyst
supplying a suitable set of initial guesses. A systematic way to generate
the set of initial guesses to use is provided by \emph{continuation}.
Suppose that the problem incorporates some set of parameters $k$,
which for the liquid crystal problem includes the elastic constants
and preferred pitch $q_{0}$. Given a set of solutions for some initial
value of these parameters $k_{0}=\bar{k}_{0}$, we use each of these
solutions as an initial guess for Newton's method at a nearby parameter
$k_{0}=\bar{k}_{0}+\delta k$, deflating each solution as we find
it. Subsequently, we use the full power of the deflation approach
to search for new solution branches that, if discovered, can be extended
to other values of $k_{0}$ using standard continuation techniques.
The solutions at $k_{0}=\bar{k}_{0}+\delta k$ can in turn be used
as initial guesses to find the solutions at $k_{0}=\bar{k}_{0}+2\delta k$,
and so on. This combination of deflation and continuation is referred
to as \emph{deflated continuation} and is an even more powerful algorithm
than standard deflation applied to a single nonlinear problem \cite{Farrell4}.
It can be interpreted physically as computing a bifurcation diagram,
a portrait of the solutions of an equation as a parameter varies.
We shall use both the deflation and the deflated continuation approach
to resolve ground state solutions of the cholesteric problem in the
next section.

\section{Results\label{sec:Cholesterics-with-Curved}}

\begin{figure}
\begin{centering}
\includegraphics[width=1\columnwidth]{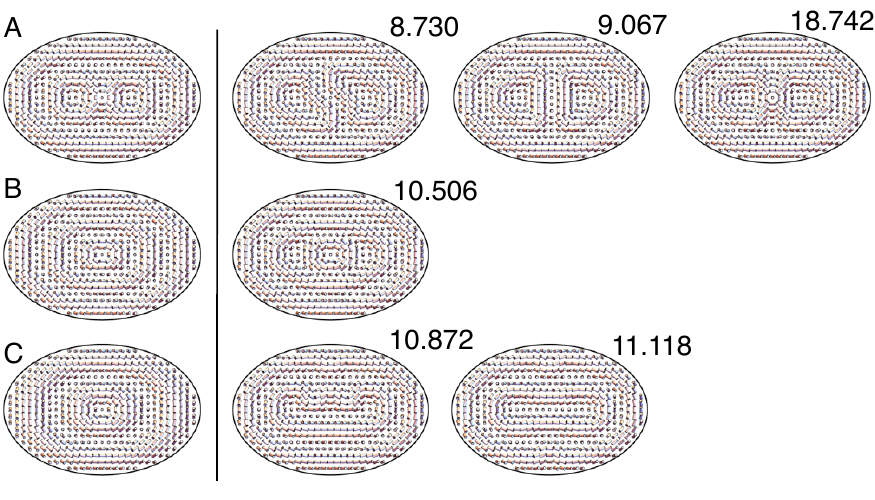} 
\par\end{centering}
\caption{\textbf{\label{fig:DeflationExample}Example solution set found with
deflation. }Solution set for cholesteric pitch $q_{0}=8$ and aspect
ratio $\mu=1.5$. Rows \textbf{A}\textemdash \textbf{C} depict different
initial guesses (left) and the solution set (right) recovered for
each through successive applications of the deflation operator (\ref{eq:deflationoperator}).
The computed free energy of each solution is also given.}
\end{figure}

\begin{figure*}
\begin{centering}
\includegraphics[width=2\columnwidth]{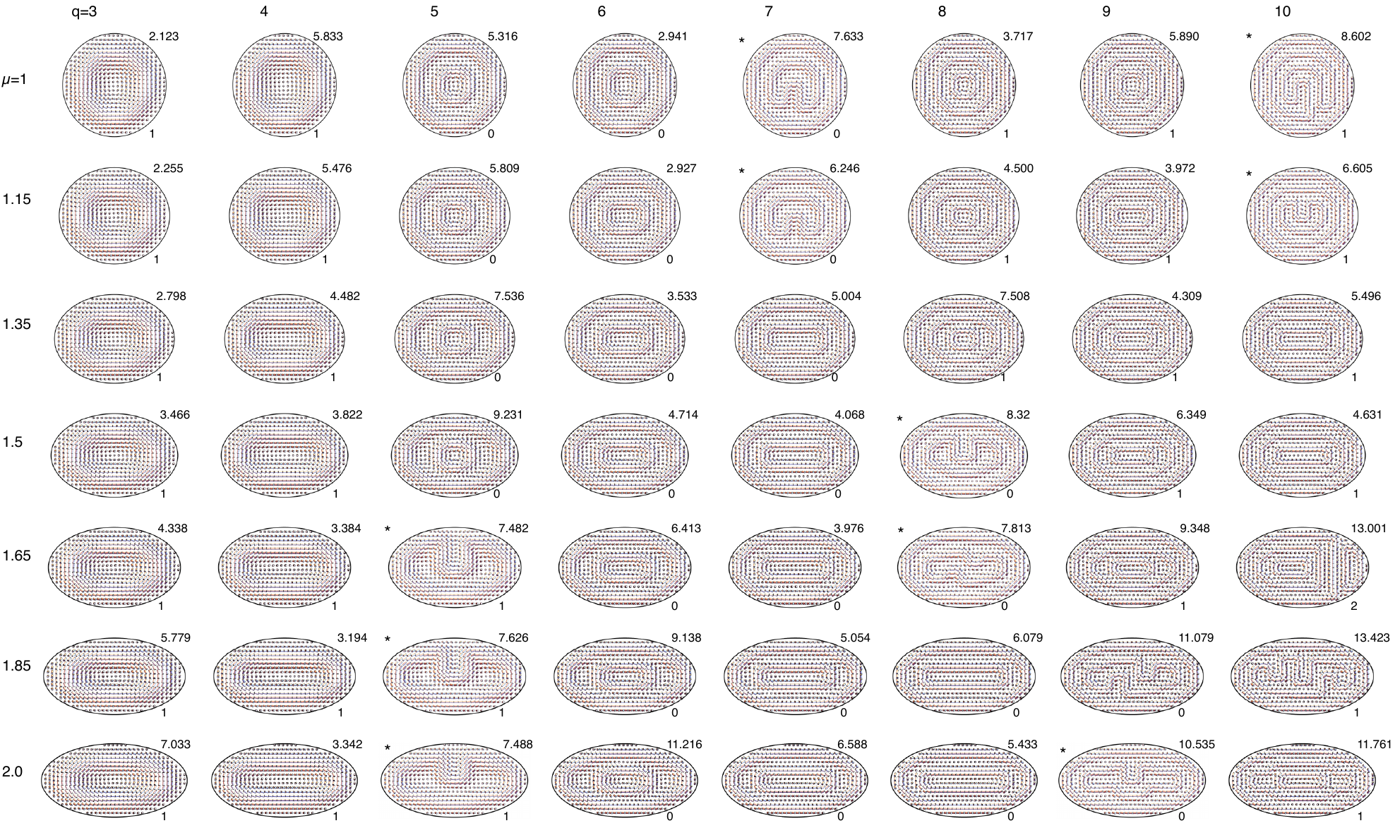} 
\par\end{centering}
\caption{\textbf{\label{fig:Cholesteric-liquid-crystal}Cholesteric liquid
crystal in an elliptical domain}. Ground state solutions shown as
a function of pitch $q_{0}$ and aspect ratio $\mu$. For each solution,
the value of the energy functional is displayed on the top right,
with the skyrmion number shown bottom right. Solutions indicated by an
asterisk on the top left were found using the deflated continuation technique. }
\end{figure*}

We apply the deflation algorithm described above to the cholesteric
problem in an ellipse, varying the aspect ratio of the domain, $\mu$,
and preferred cholesteric pitch, $q_{0}$. In each simulation, the
director is held fixed on the boundary such that the director points
out of the plane, i.e. $\mathbf{n}=(0,0,1)$. Elastic constants are
chosen to be $K_{1}=1$, $K_{2}=3.2$ and $K_{3}=1.1$, corresponding
to the exotic splay-bend cholesteric described above. The computational
domain is centered on the origin with major axis parallel to the $x$-axis.
The area of each bounding ellipse is held fixed at $\frac{3\pi}{2}$.

Our multilevel finite-element code used to compute stationary points
of the free energy (\ref{eq:Frank}) is thoroughly described elsewhere~\cite{Emerson1,Emerson2,Emerson3,adler2017combining}.
Briefly, the code uses the Cartesian representation of the director
$\mathbf{n}=(n_{x},n_{y},n_{z})$ and directly finds equilibrium points
of the Frank energy (\ref{eq:Frank}) by applying Newton linearisation
to the first-order optimality conditions in variational form, resulting
from the constrained minimisation. The code is based on deal.II \cite{BangerthHartmannKanschat2007}
and features mesh refinement and nested iteration \cite{Starke1},
such that the problem is discretised and solved first on coarse meshes
where computation is cheap, resolving major solution features, and
then interpolated to more refined meshes to determine finer attributes
of the computed approximation. Nested iteration has been shown to
significantly improve computational efficiency in a wide variety of
problems including liquid crystal simulations~\cite{Emerson3,adler2010nested,manteuffel2006first}.
Here, we use a nested iteration mesh hierarchy of four mesh levels
of refinement with $4,884$ degrees of freedom on the coarsest grid
and ending with 297,988 degrees of freedom at the finest level. Finally,
a damping factor, $\omega$, is applied to each Newton step for both
the undeflated and deflated systems. This damped Newton stepping is
combined with an increased step size, $\bar{\omega}$, when the nonlinear
residual drops below 0.1. The accelerated Newton stepping is applied
to increase the rate of convergence when a candidate solution begins
to closely satisfy the optimality conditions.

As a first example, in Figure \ref{fig:DeflationExample} we display
the results of a typical run for aspect ratio $\mu=1.5$ and $q_{0}=8$.
The algorithm is initialised from three initial guesses (shown in
the left column of Fig. \ref{fig:DeflationExample}). As anticipated,
deflation enables the discovery of several solutions for each value
by successively removing them with the deflation operator. Several
of these solutions possess energetically degenerate partners that
are obtained by simple reflection about an axis of the ellipse. These
are also found by deflation, even though knowledge of the symmetry
of the problem is not explicitly built into the algorithm.

It is important to note that the solutions to which the code converges
are \emph{stationary} points of the Lagrangian (Frank energy plus
unit-length constraint), not necessarily minimisers of the energy.
It is therefore highly desirable to characterise the nature of each
solution as it is found, i.e. determine if it is a local minimum,
a local maximum, or a saddle point. To do this, we must verify the
second-order sufficiency conditions: a stationary point is a local
minimum if the reduced Hessian of the energy (the Hessian projected
onto the nullspace of the linearised constraints, i.e.~restricted
to feasible perturbations) is positive-definite. One approach would
be to assemble the \foreignlanguage{british}{linearised} constraint
Jacobian, compute a (dense) basis for its nullspace using the singular
value decomposition, construct the (dense) reduced Hessian, and compute
its eigenvalues; however, this would be unaffordably expensive. A better
way is to exploit the fact that the eigenvalues of the reduced Hessian
of the energy are related to the eigenvalues of the (sparse) Hessian
of the Lagrangian: by counting the number of negative eigenvalues
of the Hessian of the Lagrangian, and comparing it to the dimension
of the constraint space, we can determine the inertia (the number
of positive, zero, and negative eigenvalues) of the associated reduced
Hessian of the energy \cite[Thm.~16.3]{nocedal2006}. This allows
for the characterisation of the nature of each solution found using
a single sparse $LDL^{T}$ decomposition, computed using the FEniCS,
PETSc and MUMPS software packages \cite{logg2011,balay2015,amestoy2001}.

In Figure \ref{fig:Cholesteric-liquid-crystal}, we show the computed
ground state (lowest-energy solution) as a function of $\mu$ and
$q_{0}$. For each solution, we display the value of the energy functional
and also compute the skyrmion number~\cite{bogdanov2003skyrmions},
\begin{equation}
Q=\frac{1}{4\pi}\int\mathbf{n}\cdot\left(\frac{\partial\mathbf{n}}{\partial x}\times\frac{\partial\mathbf{n}}{\partial y}\right)dA,\label{eq:skyrmion number}
\end{equation}
a topological index that represents the number of times $\mathbf{n}$
covers the unit sphere. Such indices help identify topologically distinct
solutions: as the parameters $\mu$ and $q_{0}$ are slowly varied
in Fig. \ref{fig:Cholesteric-liquid-crystal}, the ground state mostly
changes smoothly. However, between certain values, e.g. $q_{0}=4$
and $5$ with $\mu=1$, a transition to a new solution as the ground
state occurs; this is accompanied by a change in the skyrmion number.
Some solutions that are quite different in appearance e.g. $\mu=1.85$
between $q_{0}=4$ and $5$ or $q_{0}=8$ and $9$ have identical
$Q$ because they are linked by a continuous deformation of the director
field. While deflated continuation enables us to find intermediate
solutions between chosen values, and resolve transitions that
take place, the skyrmion number provides a partial classification of the distinct
branches that arise.

\begin{figure}
\includegraphics{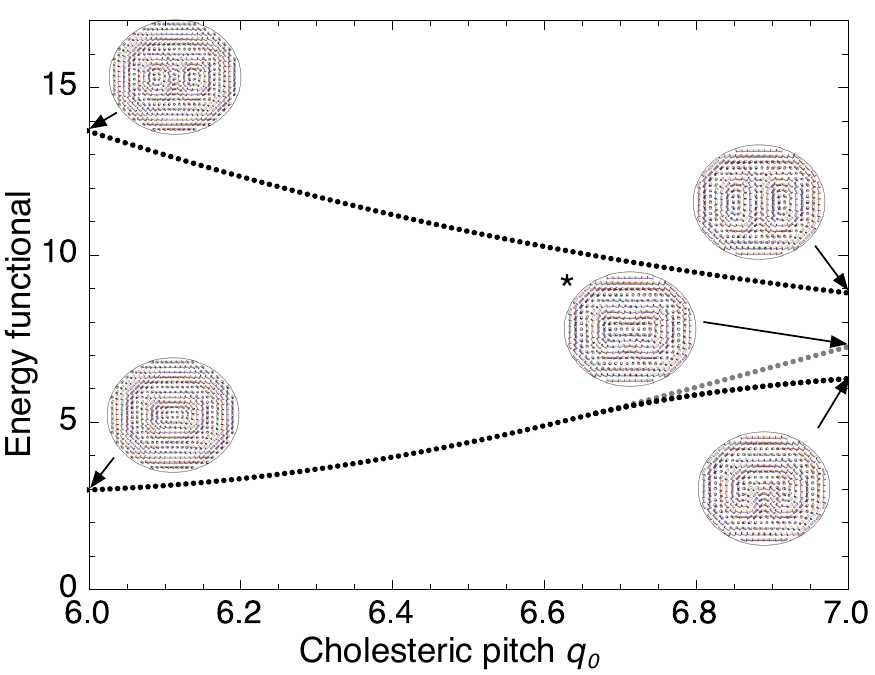}

\caption{\textbf{\label{fig:Resolve}Bifurcation diagram} as a function of
$q_{0}$ generated by deflated continuation for aspect ratio $\mu=1.15$.
The solution set is visualised at $q_{0}=6$ and $q_{0}=7$. Black
points represent stable solutions and gray points indicate one unstable
direction. The lowest energy, yet non-stable, solution identified
by deflation without continuation for $q_{0}=7$ is indicated by an
asterisk. }
\end{figure}

For certain values of $\mu$ and $q_{0}$, the solution set discovered
by deflation alone failed to include a minimal energy solution that
was stable. These values are indicated in Figure \ref{fig:Cholesteric-liquid-crystal}
with an asterisk. For these values, we applied the deflated continuation
technique described above to identify the stable ground state shown
in Figure \ref{fig:Cholesteric-liquid-crystal}. For instance, consider
the case $\mu=1.15$ and $q_{0}=7$. The lowest energy solution found
using deflation is displayed as an inset indicated by an asterisk
in Fig. \ref{fig:Resolve}, but possesses unstable directions according
to the Hessian analysis described above. We therefore use the $\mu=1.15$
and $q_{0}=6$ solution set and continue in $q_{0}$ from $6$ to
$7$. The energies of intermediate solutions obtained in the process
are plotted in Fig. \ref{fig:Resolve} as a bifurcation diagram; the
initial and final solution set recovered in this process are displayed
in Fig. \ref{fig:Resolve} as insets. A new pair of solutions, not
in the $q_{0}=6$ solution set, emerges through a fold bifurcation
at approximately $q_{0}\approx6.67$ and represents the stable ground
state at higher values of $q_{0}$; the prior ground state becomes
higher in energy and is now unstable. 

The same procedure was applied to all the problematic cases in Fig.
\ref{fig:Cholesteric-liquid-crystal}, where the solutions shown are
the lowest energy solutions found and are all verified as stable.
This example illustrates the power of deflated continuation to track
distinct branches in the solution set and identify solutions very
different from the initial guesses provided. While it remains possible
that the true ground state remains elusive for some values in $(\mu,q_{0})$
space, it is clear that deflation and deflated continuation are powerful
tools to assist in the assembly of phase diagrams.

The solutions found in Fig. \ref{fig:Cholesteric-liquid-crystal}
catalogue the interesting interplay of the elastic constants, cholesteric
parameter, and confining geometry. For $\mu=1$, a circular domain,
increasing $q_{0}$ initially leads only to the incorporation of additional
rotations as expected. Around a critical value of $q_{0}=7$, the
contours of constant orientation are greatly deformed as the number
of radial rotations in the channel increases from $\pi$ at $q_{0}=6$
to $3\pi/2$ for $q_{0}=8$. A similarly deformed structure is visible
at $q_{0}=10$, which is apparently close to a jump from $3\pi/2$
rotations to $2\pi$.

For higher aspect ratios, the contours of constant orientation can
be deformed by the geometry of the channel. For example, for aspect
ratio $\mu=1.5$ and $q_{0}\le7$, the ground state consists of the
director rotating by $2\pi$ about the radial direction from the center;
above this value an extra $\pi$ rotation is incorporated. Comparing
the shape of contours of constant tilt, notice that the interior ``layer''
is approximately circular for $q_{0}=5$, but becomes more elongated
with increasing $q_{0}$. For $q_{0}=8$, the ground state is strikingly
different: a highly deformed interior layer is accommodated
within one contiguous outer layer. The ground state for $q_{0}=9$
reverts to the expected pattern, simply incorporating an additional
twist. Furthermore, as $\mu$ increases, the transition points between
states with different amounts of twist occur at higher values of $q_{0}$,
and are typically proceeded by substantial layer deformation. Therefore,
the confining geometry plays a role in deterring or encouraging the
addition of layers.

We note that deflation uncovers a particularly large number of solutions
for $q_{0}=8$ and that many of the solutions have relatively low
energy compared to the ground state. For other values of $q_{0}$,
only a few of the solutions are close to the ground state in energy.
We speculate that this phenomenon is due to \emph{commensurability},
with some values of $q_{0}$ being more compatible with the shape
of the domain than others. For a circular domain, commensurate solutions
exist where $q_{0}$ happens to allow an integer multiple of $\pi$
rotations from the center. Maximally strained solutions occur between
these special values, potentially inducing deformation of the layers
to relieve the frustration. This is clearly visible at $\mu=1.5$
and $q_{0}=8$, or at $\mu=1.85$ and $q_{0}\geq9$, where the inner
layer in both cases is highly tortuous to fill the interior of the
domain.

\begin{figure*}
\begin{centering}
\includegraphics[width=7in]{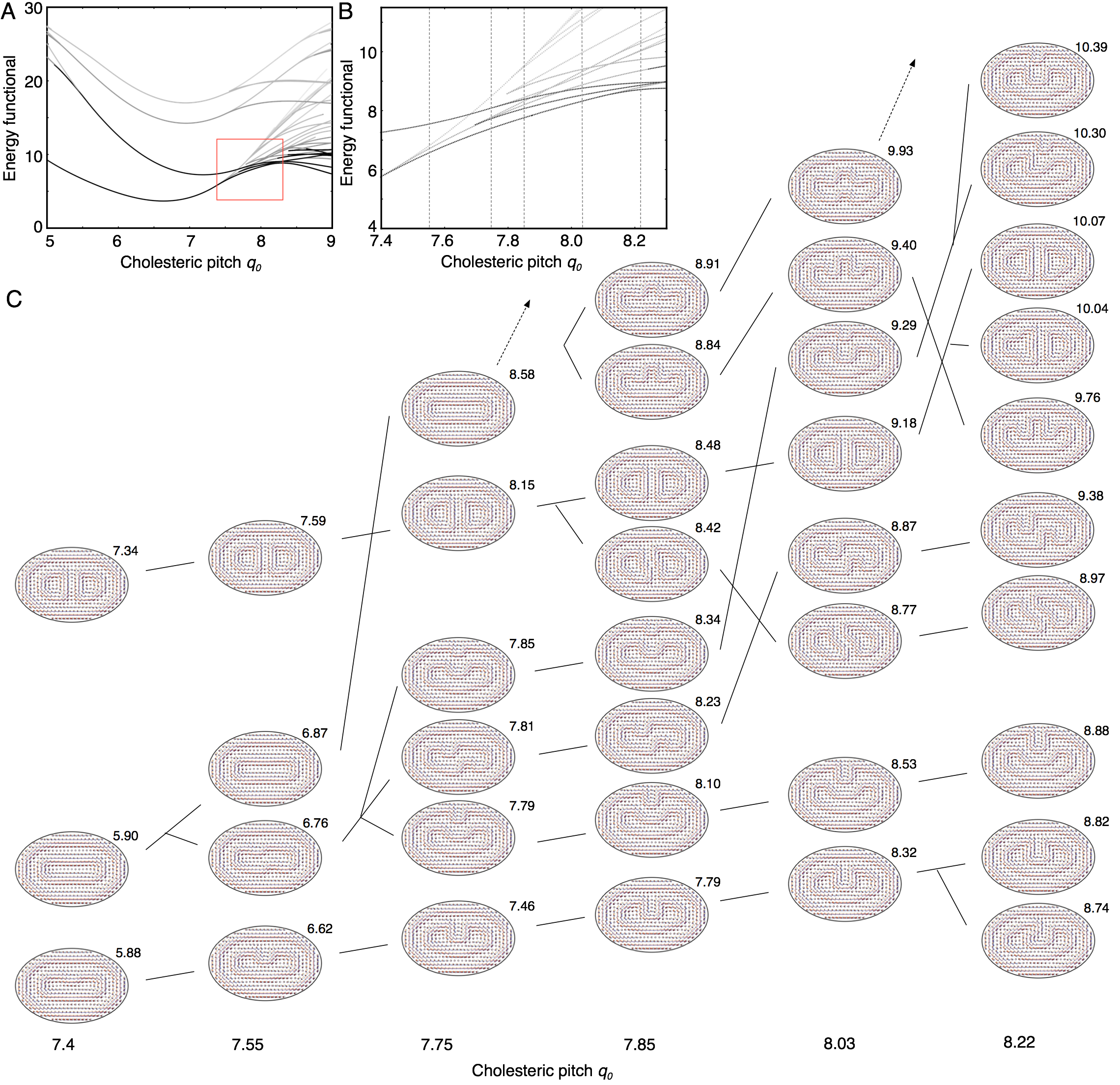} 
\par\end{centering}
\caption{\label{fig:continuation-q} \textbf{Deflated continuation. A }Bifurcation
diagram computed for fixed $\mu=1.5$ and continuing in $q_{0}$.
Points are colored by the number of unstable directions, with black
indicating a stable solution, and lighter grey indicating more unstable
directions. A particularly dense portion of the diagram, outlined
in red, is shown in greater detail in \textbf{B }where vertical dashed
lines indicate values of $q_{0}$ for which the solution set is displayed
in \textbf{C}. }
\end{figure*}

To resolve the sequence of transitions that occurs around one of the
maximally strained solutions, we visualise a bifurcation diagram in
Figure \ref{fig:continuation-q} for an elliptical domain with aspect
ratio $\mu=1.5$ and with the preferred pitch ranging from $q_{0}=5$
to $q_{0}=9$. We \foreignlanguage{british}{initialise} the computation
with the solutions found by our previous analysis. The diagram shown
in \ref{fig:continuation-q}A displays a relatively small solution
set for $q_{0}<7$, but above $q_{0}\approx7.4$, a dense thicket
of additional solutions appears. As is visible in the expanded region
depicted in \ref{fig:continuation-q}B, this consists of a rapid succession
of fold and pitchfork bifurcations. Moreover, the region between $q_0 = 7.4$ 
and $9.0$ contains a notable number of stable solutions with free energies 
in relatively close proximity to the ground state. The corresponding solutions at
several values of $q_{0}$ are displayed in \ref{fig:continuation-q}C,
illustrating the striking complexity of solution sets that can be
uncovered using deflated continuation.

\begin{figure}
\includegraphics{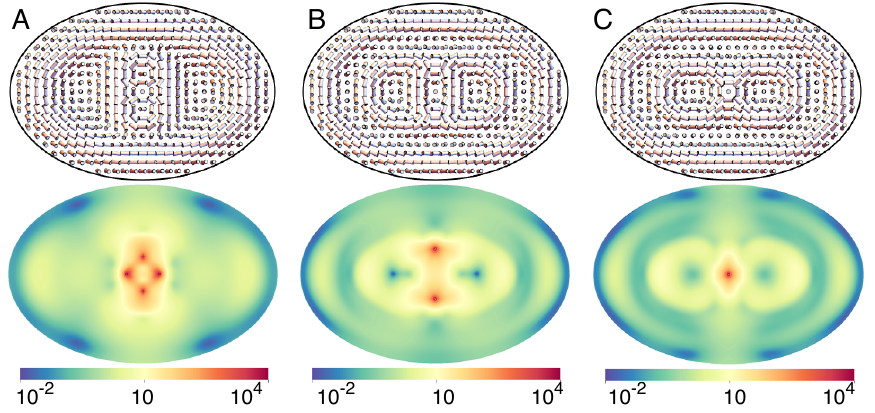}

\caption{\textbf{\label{fig:Channel-Induced-Disclinations:}Channel Induced
Disclinations: }(Above) Plots of the director field and (below) the
free energy density, displaying the formation of defects for different
cholesteric pitch $q_{0}$. Regions of high energy density in red
indicate the location of each defect. Each ellipse has an aspect ratio
of $\mu=1.5$. Cholesteric configurations with \textbf{A} $q_{0}=5.0$
forming four symmetrically arranged defects with free energy of $67.684$;
\textbf{B} $q_{0}=7.0$ with two disclinations and free energy of
$32.434$; \textbf{C $q_{0}=7.0$, }free energy $14.959$,\textbf{
}and a single central defect.}
\end{figure}

Because deflation suppresses solutions close to the initial guess,
even very distant solutions can be recovered. A particularly important
result is that, while our initial guesses are smooth functions, the
algorithm was able to spontaneously identify solutions with disclinations.
Confined cholesterics in rectangular domains or channels have been
experimentally and numerically shown to exhibit structures with disclinations~\cite{YHKim1,Ackerman1,Guo1},
as discussed in the introduction. The existence, type, and number
of the disclinations were shown to be modulated by changes to the
depth and width of the channel, as well as the cholesteric pitch.
Figure \ref{fig:Channel-Induced-Disclinations:} displays three examples
for $\mu=1.5$ and varying $q_{0}$ with computed director fields
and associated elastic energy densities from which disclinations are
readily identified. For $q_{0}=5$, the solution shown in Fig. \ref{fig:Channel-Induced-Disclinations:}A
has four defect points, arranged in a diamond pattern near the center
of the domain. The solutions in Fig. \ref{fig:Channel-Induced-Disclinations:}B
and C were found for $q_{0}=7$ and possess two and one disclinations,
respectively. Comparing the solutions' free energies, it is clear
that the single defect structure is energetically preferred. We note
that the energy of the solution in \ref{fig:Channel-Induced-Disclinations:}C
corresponds to the third lowest energy (and first unstable state)
shown for $q_{0}=7$ in the bifurcation diagram in Figure \ref{fig:continuation-q}A.
Thus, our results suggest that the propensity of the cholesteric to
forming metastable structures with disclinations in elliptical channels,
perhaps upon quenching from the isotropic phase, strongly depends
on the aspect ratio of the channel boundary. Moreover, our numerical
experiments indicate that multiple equilibrium configurations with
distinct disclination patterns may exist for the same geometry and
material parameters.

The deflation technique plays a central role in the discovery of these
disclination arrangements. For instance, numerical simulations in
\cite{Guo1} relied on \emph{a priori} knowledge, gained from experimental
observations, that disclination structures should be present in order
to initialise the Newton iterations within a basin of attraction.
We emphasise that here the simulations are initialised with smooth
director fields. Multiple solutions and the emergence of disclinations
occurred as spontaneous discoveries enabled by the deflation computations.
In situations where experimental and analytical information is limited,
such numerical capabilities facilitate a more robust and thorough
exploration of the admissible solution space of a given problem.

\section{Conclusion\label{sec:Conclusion}}

In this paper, we present a new technique, deflation, for recovering
equilibrium solutions of the free energy of a liquid crystal. The
utility of the method is shown on a toy example and then used to determine
the structure of a cholesteric in an elliptical domain. The ground
state is identified for a range of aspect ratios $\mu$ and preferred
pitches $q_{0}$, showing gradual deformation of the solutions
as a function of these parameters and transitions to different solutions
at critical values. For selected values of $\mu$ and $q_{0}$, we
compute the bifurcation diagram, finding remarkably dense solution sets near the transition points.
In future work, we will apply the method to characterise the solution
set of more complex geometries involving cholesterics, such as the
rich 3D structures observed in \cite{smalyukh2010three}. 

The deflation methodology significantly enhances the utility of Newton
iterations applied to nonlinear systems by enabling Newton's method
to converge to multiple solutions from the same initial guess. Applying
the deflation operator ensures that subsequent Newton iterations do
not converge to previously discovered solutions and effectively modifies
the basin of attraction to include unknown solutions. While convergence
to multiple solutions is not guaranteed through use of the deflation
operator, sufficient conditions for convergence of deflated iterations
are constructed in \cite{Farrell4} based on a generalisation of the
Rall-Rheinboldt theorem. Our results illuminate the effect of deflation,
but also highlight the case that the use of multiple initial guesses
to discover all solutions is not entirely eliminated with deflation.
However, deflated continuation systematically provides a sequence
of good initial guesses and increases overall efficiency and reliability
of multiple solution discovery through more systematic exploration
of the solution space. 

More generally, deflation and deflated continuation therefore allow
theorists to recover energetically low-lying solutions in which a
liquid crystal may become kinetically trapped. It can also be used
to track different solution branches when a system exhibits a bifurcation
with respect to some external parameter, e.g. the applied field in
a Freedericksz transition. The method is very general and can be readily
adapted to different representations, e.g. the Q-tensor, and parametrisations;
it is likely to be most useful in systems where little analytical
guidance or experimental imaging are available.

\begin{acknowledgments}
The work of SPM was partially supported by an NSERC discovery grant.
TJA is supported by a Cottrell Award from the Research Corporation
for Science Advancement, and a CAREER award from the National Science
Foundation (award number DMR-CMMT-1654283). PEF is supported by an
EPSRC Early Career Research Fellowship (award number EP/K030930/1).
The authors contributed to the work as follows: SPM devised the toy
example in Section III; DBE, PEF, and JHA performed the simulations
in Section IV and all authors contributed to their analysis; TJA wrote
the paper, which was revised by all authors. 
\end{acknowledgments}

\appendix
%


\begin{thebibliography}{39}%
\makeatletter
\providecommand \@ifxundefined [1]{%
 \@ifx{#1\undefined}
}%
\providecommand \@ifnum [1]{%
 \ifnum #1\expandafter \@firstoftwo
 \else \expandafter \@secondoftwo
 \fi
}%
\providecommand \@ifx [1]{%
 \ifx #1\expandafter \@firstoftwo
 \else \expandafter \@secondoftwo
 \fi
}%
\providecommand \natexlab [1]{#1}%
\providecommand \enquote  [1]{``#1''}%
\providecommand \bibnamefont  [1]{#1}%
\providecommand \bibfnamefont [1]{#1}%
\providecommand \citenamefont [1]{#1}%
\providecommand \href@noop [0]{\@secondoftwo}%
\providecommand \href [0]{\begingroup \@sanitize@url \@href}%
\providecommand \@href[1]{\@@startlink{#1}\@@href}%
\providecommand \@@href[1]{\endgroup#1\@@endlink}%
\providecommand \@sanitize@url [0]{\catcode `\\12\catcode `\$12\catcode
  `\&12\catcode `\#12\catcode `\^12\catcode `\_12\catcode `\%12\relax}%
\providecommand \@@startlink[1]{}%
\providecommand \@@endlink[0]{}%
\providecommand \url  [0]{\begingroup\@sanitize@url \@url }%
\providecommand \@url [1]{\endgroup\@href {#1}{\urlprefix }}%
\providecommand \urlprefix  [0]{URL }%
\providecommand \Eprint [0]{\href }%
\providecommand \doibase [0]{http://dx.doi.org/}%
\providecommand \selectlanguage [0]{\@gobble}%
\providecommand \bibinfo  [0]{\@secondoftwo}%
\providecommand \bibfield  [0]{\@secondoftwo}%
\providecommand \translation [1]{[#1]}%
\providecommand \BibitemOpen [0]{}%
\providecommand \bibitemStop [0]{}%
\providecommand \bibitemNoStop [0]{.\EOS\space}%
\providecommand \EOS [0]{\spacefactor3000\relax}%
\providecommand \BibitemShut  [1]{\csname bibitem#1\endcsname}%
\let\auto@bib@innerbib\@empty
\bibitem [{\citenamefont {Bahr}\ and\ \citenamefont
  {Kitzerow}(2001)}]{bahr2001chirality}%
  \BibitemOpen
  \bibfield  {author} {\bibinfo {author} {\bibfnamefont {C.}~\bibnamefont
  {Bahr}}\ and\ \bibinfo {author} {\bibfnamefont {H.-S.}\ \bibnamefont
  {Kitzerow}},\ }\href@noop {} {\emph {\bibinfo {title} {Chirality in liquid
  crystals}}}\ (\bibinfo  {publisher} {Springer},\ \bibinfo {year}
  {2001})\BibitemShut {NoStop}%
\bibitem [{\citenamefont {Bogdanov}\ \emph {et~al.}(2003)\citenamefont
  {Bogdanov}, \citenamefont {R{\"o}{\ss}ler},\ and\ \citenamefont
  {Shestakov}}]{bogdanov2003skyrmions}%
  \BibitemOpen
  \bibfield  {author} {\bibinfo {author} {\bibfnamefont {A.}~\bibnamefont
  {Bogdanov}}, \bibinfo {author} {\bibfnamefont {U.}~\bibnamefont
  {R{\"o}{\ss}ler}}, \ and\ \bibinfo {author} {\bibfnamefont {A.}~\bibnamefont
  {Shestakov}},\ }\href@noop {} {\bibfield  {journal} {\bibinfo  {journal}
  {Phys. Rev. E}\ }\textbf {\bibinfo {volume} {67}},\ \bibinfo {pages} {016602}
  (\bibinfo {year} {2003})}\BibitemShut {NoStop}%
\bibitem [{\citenamefont {Smalyukh}\ \emph {et~al.}(2010)\citenamefont
  {Smalyukh}, \citenamefont {Lansac}, \citenamefont {Clark},\ and\
  \citenamefont {Trivedi}}]{smalyukh2010three}%
  \BibitemOpen
  \bibfield  {author} {\bibinfo {author} {\bibfnamefont {I.~I.}\ \bibnamefont
  {Smalyukh}}, \bibinfo {author} {\bibfnamefont {Y.}~\bibnamefont {Lansac}},
  \bibinfo {author} {\bibfnamefont {N.~A.}\ \bibnamefont {Clark}}, \ and\
  \bibinfo {author} {\bibfnamefont {R.~P.}\ \bibnamefont {Trivedi}},\
  }\href@noop {} {\bibfield  {journal} {\bibinfo  {journal} {Nature Materials}\
  }\textbf {\bibinfo {volume} {9}},\ \bibinfo {pages} {139} (\bibinfo {year}
  {2010})}\BibitemShut {NoStop}%
\bibitem [{\citenamefont {Fukuda}\ and\ \citenamefont
  {{\v{Z}}umer}(2011)}]{fukuda2011quasi}%
  \BibitemOpen
  \bibfield  {author} {\bibinfo {author} {\bibfnamefont {J.-i.}\ \bibnamefont
  {Fukuda}}\ and\ \bibinfo {author} {\bibfnamefont {S.}~\bibnamefont
  {{\v{Z}}umer}},\ }\href@noop {} {\bibfield  {journal} {\bibinfo  {journal}
  {Nature Communications}\ }\textbf {\bibinfo {volume} {2}},\ \bibinfo {pages}
  {246} (\bibinfo {year} {2011})}\BibitemShut {NoStop}%
\bibitem [{\citenamefont {Ackerman}\ \emph
  {et~al.}(2014{\natexlab{a}})\citenamefont {Ackerman}, \citenamefont
  {Trivedi}, \citenamefont {Senyuk}, \citenamefont {van~de Lagemaat},\ and\
  \citenamefont {Smalyukh}}]{ackerman2014two}%
  \BibitemOpen
  \bibfield  {author} {\bibinfo {author} {\bibfnamefont {P.~J.}\ \bibnamefont
  {Ackerman}}, \bibinfo {author} {\bibfnamefont {R.~P.}\ \bibnamefont
  {Trivedi}}, \bibinfo {author} {\bibfnamefont {B.}~\bibnamefont {Senyuk}},
  \bibinfo {author} {\bibfnamefont {J.}~\bibnamefont {van~de Lagemaat}}, \ and\
  \bibinfo {author} {\bibfnamefont {I.~I.}\ \bibnamefont {Smalyukh}},\
  }\href@noop {} {\bibfield  {journal} {\bibinfo  {journal} {Phys. Rev. E}\
  }\textbf {\bibinfo {volume} {90}},\ \bibinfo {pages} {012505} (\bibinfo
  {year} {2014}{\natexlab{a}})}\BibitemShut {NoStop}%
\bibitem [{\citenamefont {Leonov}\ \emph {et~al.}(2014)\citenamefont {Leonov},
  \citenamefont {Dragunov}, \citenamefont {R{\"o}{\ss}ler},\ and\ \citenamefont
  {Bogdanov}}]{leonov2014theory}%
  \BibitemOpen
  \bibfield  {author} {\bibinfo {author} {\bibfnamefont {A.}~\bibnamefont
  {Leonov}}, \bibinfo {author} {\bibfnamefont {I.}~\bibnamefont {Dragunov}},
  \bibinfo {author} {\bibfnamefont {U.}~\bibnamefont {R{\"o}{\ss}ler}}, \ and\
  \bibinfo {author} {\bibfnamefont {A.}~\bibnamefont {Bogdanov}},\ }\href@noop
  {} {\bibfield  {journal} {\bibinfo  {journal} {Phys. Rev. E}\ }\textbf
  {\bibinfo {volume} {90}},\ \bibinfo {pages} {042502} (\bibinfo {year}
  {2014})}\BibitemShut {NoStop}%
\bibitem [{\citenamefont {Ackerman}\ \emph {et~al.}(2015)\citenamefont
  {Ackerman}, \citenamefont {van~de Lagemaat},\ and\ \citenamefont
  {Smalyukh}}]{ackerman2015self}%
  \BibitemOpen
  \bibfield  {author} {\bibinfo {author} {\bibfnamefont {P.~J.}\ \bibnamefont
  {Ackerman}}, \bibinfo {author} {\bibfnamefont {J.}~\bibnamefont {van~de
  Lagemaat}}, \ and\ \bibinfo {author} {\bibfnamefont {I.~I.}\ \bibnamefont
  {Smalyukh}},\ }\href@noop {} {\bibfield  {journal} {\bibinfo  {journal}
  {Nature Communications}\ }\textbf {\bibinfo {volume} {6}} (\bibinfo {year}
  {2015})}\BibitemShut {NoStop}%
\bibitem [{\citenamefont {Ackerman}\ and\ \citenamefont
  {Smalyukh}(2016)}]{ackerman2016reversal}%
  \BibitemOpen
  \bibfield  {author} {\bibinfo {author} {\bibfnamefont {P.~J.}\ \bibnamefont
  {Ackerman}}\ and\ \bibinfo {author} {\bibfnamefont {I.~I.}\ \bibnamefont
  {Smalyukh}},\ }\href@noop {} {\bibfield  {journal} {\bibinfo  {journal}
  {Phys. Rev. E}\ }\textbf {\bibinfo {volume} {93}},\ \bibinfo {pages} {052702}
  (\bibinfo {year} {2016})}\BibitemShut {NoStop}%
\bibitem [{\citenamefont {Chen}\ \emph {et~al.}(2013)\citenamefont {Chen},
  \citenamefont {Ackerman}, \citenamefont {Alexander}, \citenamefont {Kamien},\
  and\ \citenamefont {Smalyukh}}]{chen2013generating}%
  \BibitemOpen
  \bibfield  {author} {\bibinfo {author} {\bibfnamefont {B.~G.-g.}\
  \bibnamefont {Chen}}, \bibinfo {author} {\bibfnamefont {P.~J.}\ \bibnamefont
  {Ackerman}}, \bibinfo {author} {\bibfnamefont {G.~P.}\ \bibnamefont
  {Alexander}}, \bibinfo {author} {\bibfnamefont {R.~D.}\ \bibnamefont
  {Kamien}}, \ and\ \bibinfo {author} {\bibfnamefont {I.~I.}\ \bibnamefont
  {Smalyukh}},\ }\href@noop {} {\bibfield  {journal} {\bibinfo  {journal}
  {Phys. Rev. Lett.}\ }\textbf {\bibinfo {volume} {110}},\ \bibinfo {pages}
  {237801} (\bibinfo {year} {2013})}\BibitemShut {NoStop}%
\bibitem [{\citenamefont {Cattaneo}\ \emph {et~al.}(2016)\citenamefont
  {Cattaneo}, \citenamefont {Kos}, \citenamefont {Savoini}, \citenamefont
  {Kouwer}, \citenamefont {Rowan}, \citenamefont {Ravnik}, \citenamefont
  {Mu{\v{s}}evi{\v{c}}},\ and\ \citenamefont {Rasing}}]{cattaneo2016electric}%
  \BibitemOpen
  \bibfield  {author} {\bibinfo {author} {\bibfnamefont {L.}~\bibnamefont
  {Cattaneo}}, \bibinfo {author} {\bibfnamefont {{\v{Z}}.}~\bibnamefont {Kos}},
  \bibinfo {author} {\bibfnamefont {M.}~\bibnamefont {Savoini}}, \bibinfo
  {author} {\bibfnamefont {P.}~\bibnamefont {Kouwer}}, \bibinfo {author}
  {\bibfnamefont {A.}~\bibnamefont {Rowan}}, \bibinfo {author} {\bibfnamefont
  {M.}~\bibnamefont {Ravnik}}, \bibinfo {author} {\bibfnamefont
  {I.}~\bibnamefont {Mu{\v{s}}evi{\v{c}}}}, \ and\ \bibinfo {author}
  {\bibfnamefont {T.}~\bibnamefont {Rasing}},\ }\href@noop {} {\bibfield
  {journal} {\bibinfo  {journal} {Soft Matter}\ }\textbf {\bibinfo {volume}
  {12}},\ \bibinfo {pages} {853} (\bibinfo {year} {2016})}\BibitemShut
  {NoStop}%
\bibitem [{\citenamefont {Kim}\ \emph {et~al.}(2015{\natexlab{a}})\citenamefont
  {Kim}, \citenamefont {Gim}, \citenamefont {Jung},\ and\ \citenamefont
  {Yoon}}]{kim2015periodic}%
  \BibitemOpen
  \bibfield  {author} {\bibinfo {author} {\bibfnamefont {Y.~H.}\ \bibnamefont
  {Kim}}, \bibinfo {author} {\bibfnamefont {M.-J.}\ \bibnamefont {Gim}},
  \bibinfo {author} {\bibfnamefont {H.-T.}\ \bibnamefont {Jung}}, \ and\
  \bibinfo {author} {\bibfnamefont {D.~K.}\ \bibnamefont {Yoon}},\ }\href@noop
  {} {\bibfield  {journal} {\bibinfo  {journal} {RSC Advances}\ }\textbf
  {\bibinfo {volume} {5}},\ \bibinfo {pages} {19279} (\bibinfo {year}
  {2015}{\natexlab{a}})}\BibitemShut {NoStop}%
\bibitem [{\citenamefont {Guo}\ \emph {et~al.}(2016{\natexlab{a}})\citenamefont
  {Guo}, \citenamefont {Afghah}, \citenamefont {Xiang}, \citenamefont
  {Lavrentovich}, \citenamefont {Selinger},\ and\ \citenamefont
  {Wei}}]{guo2016cholesteric}%
  \BibitemOpen
  \bibfield  {author} {\bibinfo {author} {\bibfnamefont {Y.}~\bibnamefont
  {Guo}}, \bibinfo {author} {\bibfnamefont {S.}~\bibnamefont {Afghah}},
  \bibinfo {author} {\bibfnamefont {J.}~\bibnamefont {Xiang}}, \bibinfo
  {author} {\bibfnamefont {O.~D.}\ \bibnamefont {Lavrentovich}}, \bibinfo
  {author} {\bibfnamefont {R.~L.}\ \bibnamefont {Selinger}}, \ and\ \bibinfo
  {author} {\bibfnamefont {Q.-H.}\ \bibnamefont {Wei}},\ }\href@noop {}
  {\bibfield  {journal} {\bibinfo  {journal} {Soft Matter}\ }\textbf {\bibinfo
  {volume} {12}},\ \bibinfo {pages} {6312} (\bibinfo {year}
  {2016}{\natexlab{a}})}\BibitemShut {NoStop}%
\bibitem [{\citenamefont {Orlova}\ \emph {et~al.}(2015)\citenamefont {Orlova},
  \citenamefont {A{\ss}hoff}, \citenamefont {Yamaguchi}, \citenamefont
  {Katsonis},\ and\ \citenamefont {Brasselet}}]{orlova2015creation}%
  \BibitemOpen
  \bibfield  {author} {\bibinfo {author} {\bibfnamefont {T.}~\bibnamefont
  {Orlova}}, \bibinfo {author} {\bibfnamefont {S.~J.}\ \bibnamefont
  {A{\ss}hoff}}, \bibinfo {author} {\bibfnamefont {T.}~\bibnamefont
  {Yamaguchi}}, \bibinfo {author} {\bibfnamefont {N.}~\bibnamefont {Katsonis}},
  \ and\ \bibinfo {author} {\bibfnamefont {E.}~\bibnamefont {Brasselet}},\
  }\href@noop {} {\bibfield  {journal} {\bibinfo  {journal} {Nature
  Communications}\ }\textbf {\bibinfo {volume} {6}} (\bibinfo {year}
  {2015})}\BibitemShut {NoStop}%
\bibitem [{\citenamefont {Mart\'{i}}(2003)}]{Marti1}%
  \BibitemOpen
  \bibfield  {author} {\bibinfo {author} {\bibfnamefont {R.}~\bibnamefont
  {Mart\'{i}}},\ }in\ \href@noop {} {\emph {\bibinfo {booktitle} {Handbook in
  Metaheuristics}}},\ \bibinfo {series} {International Series in Operations
  Research \& Management Science}, Vol.~\bibinfo {volume} {57},\ \bibinfo
  {editor} {edited by\ \bibinfo {editor} {\bibfnamefont {F.}~\bibnamefont
  {Glover}}\ and\ \bibinfo {editor} {\bibfnamefont {G.~A.}\ \bibnamefont
  {Kochenberger}}}\ (\bibinfo  {publisher} {Springer},\ \bibinfo {year}
  {2003})\ pp.\ \bibinfo {pages} {355--368}\BibitemShut {NoStop}%
\bibitem [{\citenamefont {Chao}\ \emph {et~al.}(1975)\citenamefont {Chao},
  \citenamefont {Liu},\ and\ \citenamefont {Pan}}]{Chao1}%
  \BibitemOpen
  \bibfield  {author} {\bibinfo {author} {\bibfnamefont {K.-S.}\ \bibnamefont
  {Chao}}, \bibinfo {author} {\bibfnamefont {D.-K.}\ \bibnamefont {Liu}}, \
  and\ \bibinfo {author} {\bibfnamefont {C.-T.}\ \bibnamefont {Pan}},\
  }\href@noop {} {\bibfield  {journal} {\bibinfo  {journal} {IEEE Trans.
  Circuits and Syst.}\ }\textbf {\bibinfo {volume} {22}},\ \bibinfo {pages}
  {748} (\bibinfo {year} {1975})}\BibitemShut {NoStop}%
\bibitem [{\citenamefont {Allgower}\ and\ \citenamefont
  {Georg}(1993)}]{Allgower1}%
  \BibitemOpen
  \bibfield  {author} {\bibinfo {author} {\bibfnamefont {E.~L.}\ \bibnamefont
  {Allgower}}\ and\ \bibinfo {author} {\bibfnamefont {K.}~\bibnamefont
  {Georg}},\ }\href@noop {} {\bibfield  {journal} {\bibinfo  {journal} {Acta
  Numer.}\ }\textbf {\bibinfo {volume} {2}},\ \bibinfo {pages} {1} (\bibinfo
  {year} {1993})}\BibitemShut {NoStop}%
\bibitem [{\citenamefont {Farrell}\ \emph {et~al.}(2016)\citenamefont
  {Farrell}, \citenamefont {Beentjes},\ and\ \citenamefont
  {Birkisson}}]{Farrell4}%
  \BibitemOpen
  \bibfield  {author} {\bibinfo {author} {\bibfnamefont {P.~E.}\ \bibnamefont
  {Farrell}}, \bibinfo {author} {\bibfnamefont {C.~H.~L.}\ \bibnamefont
  {Beentjes}}, \ and\ \bibinfo {author} {\bibfnamefont {A.}~\bibnamefont
  {Birkisson}},\ }\href@noop {} {\bibfield  {journal} {\bibinfo  {journal}
  {Submitted, preprint available on arXiv/1603.00809}\ } (\bibinfo {year}
  {2016})}\BibitemShut {NoStop}%
\bibitem [{\citenamefont {Davidenko}(1953)}]{Davidenko1}%
  \BibitemOpen
  \bibfield  {author} {\bibinfo {author} {\bibfnamefont {D.~F.}\ \bibnamefont
  {Davidenko}},\ }\href@noop {} {\bibfield  {journal} {\bibinfo  {journal}
  {Doklady Akademii Nauk SSSR}\ }\textbf {\bibinfo {volume} {88}},\ \bibinfo
  {pages} {601} (\bibinfo {year} {1953})}\BibitemShut {NoStop}%
\bibitem [{\citenamefont {Branin}(1972)}]{Branin1}%
  \BibitemOpen
  \bibfield  {author} {\bibinfo {author} {\bibfnamefont {F.~H.}\ \bibnamefont
  {Branin}},\ }\href@noop {} {\bibfield  {journal} {\bibinfo  {journal} {IBM J.
  Res. Dev.}\ }\textbf {\bibinfo {volume} {16}},\ \bibinfo {pages} {504}
  (\bibinfo {year} {1972})}\BibitemShut {NoStop}%
\bibitem [{\citenamefont {Farrell}\ \emph {et~al.}(2015)\citenamefont
  {Farrell}, \citenamefont {Birkisson},\ and\ \citenamefont
  {Funke}}]{Farrell1}%
  \BibitemOpen
  \bibfield  {author} {\bibinfo {author} {\bibfnamefont {P.~E.}\ \bibnamefont
  {Farrell}}, \bibinfo {author} {\bibfnamefont {A.}~\bibnamefont {Birkisson}},
  \ and\ \bibinfo {author} {\bibfnamefont {S.~W.}\ \bibnamefont {Funke}},\
  }\href@noop {} {\bibfield  {journal} {\bibinfo  {journal} {SIAM J. Sci.
  Comput.}\ }\textbf {\bibinfo {volume} {37}},\ \bibinfo {pages} {A2026}
  (\bibinfo {year} {2015})}\BibitemShut {NoStop}%
\bibitem [{\citenamefont {Adler}\ \emph {et~al.}(2017)\citenamefont {Adler},
  \citenamefont {Emerson}, \citenamefont {Farrell},\ and\ \citenamefont
  {MacLachlan}}]{adler2017combining}%
  \BibitemOpen
  \bibfield  {author} {\bibinfo {author} {\bibfnamefont {J.~H.}\ \bibnamefont
  {Adler}}, \bibinfo {author} {\bibfnamefont {D.~B.}\ \bibnamefont {Emerson}},
  \bibinfo {author} {\bibfnamefont {P.~E.}\ \bibnamefont {Farrell}}, \ and\
  \bibinfo {author} {\bibfnamefont {S.~P.}\ \bibnamefont {MacLachlan}},\
  }\href@noop {} {\bibfield  {journal} {\bibinfo  {journal} {SIAM J. Sci.
  Comput.}\ }\textbf {\bibinfo {volume} {39}},\ \bibinfo {pages} {B29}
  (\bibinfo {year} {2017})}\BibitemShut {NoStop}%
\bibitem [{\citenamefont {Chapman}\ and\ \citenamefont
  {Farrell}(2017)}]{Farrell5}%
  \BibitemOpen
  \bibfield  {author} {\bibinfo {author} {\bibfnamefont {S.~J.}\ \bibnamefont
  {Chapman}}\ and\ \bibinfo {author} {\bibfnamefont {P.~E.}\ \bibnamefont
  {Farrell}},\ }\href@noop {} {\bibfield  {journal} {\bibinfo  {journal} {SIAM
  Journal on Applied Mathematics}\ } (\bibinfo {year} {2017})},\ \bibinfo
  {note} {arXiv:1609.08842 [math.CA]}\BibitemShut {NoStop}%
\bibitem [{\citenamefont {Charalampidis}\ \emph {et~al.}(2017)\citenamefont
  {Charalampidis}, \citenamefont {Kevrekidis},\ and\ \citenamefont
  {Farrell}}]{Farrell6}%
  \BibitemOpen
  \bibfield  {author} {\bibinfo {author} {\bibfnamefont {E.~G.}\ \bibnamefont
  {Charalampidis}}, \bibinfo {author} {\bibfnamefont {P.~G.}\ \bibnamefont
  {Kevrekidis}}, \ and\ \bibinfo {author} {\bibfnamefont {P.~E.}\ \bibnamefont
  {Farrell}},\ }\href@noop {} {\  (\bibinfo {year} {2017})},\ \bibinfo {note}
  {arXiv:1612.08145 [nlin.PS]}\BibitemShut {NoStop}%
\bibitem [{\citenamefont {Beller}\ \emph {et~al.}(2014)\citenamefont {Beller},
  \citenamefont {Machon}, \citenamefont {{\v{C}}opar}, \citenamefont {Sussman},
  \citenamefont {Alexander}, \citenamefont {Kamien},\ and\ \citenamefont
  {Mosna}}]{beller2014geometry}%
  \BibitemOpen
  \bibfield  {author} {\bibinfo {author} {\bibfnamefont {D.~A.}\ \bibnamefont
  {Beller}}, \bibinfo {author} {\bibfnamefont {T.}~\bibnamefont {Machon}},
  \bibinfo {author} {\bibfnamefont {S.}~\bibnamefont {{\v{C}}opar}}, \bibinfo
  {author} {\bibfnamefont {D.~M.}\ \bibnamefont {Sussman}}, \bibinfo {author}
  {\bibfnamefont {G.~P.}\ \bibnamefont {Alexander}}, \bibinfo {author}
  {\bibfnamefont {R.~D.}\ \bibnamefont {Kamien}}, \ and\ \bibinfo {author}
  {\bibfnamefont {R.~A.}\ \bibnamefont {Mosna}},\ }\href@noop {} {\bibfield
  {journal} {\bibinfo  {journal} {Physical Review X}\ }\textbf {\bibinfo
  {volume} {4}},\ \bibinfo {pages} {031050} (\bibinfo {year}
  {2014})}\BibitemShut {NoStop}%
\bibitem [{\citenamefont {Le}\ \emph {et~al.}(2015)\citenamefont {Le},
  \citenamefont {Aya}, \citenamefont {Ogino}, \citenamefont {Okano},
  \citenamefont {Araoka},\ and\ \citenamefont {Takezoe}}]{le2015large}%
  \BibitemOpen
  \bibfield  {author} {\bibinfo {author} {\bibfnamefont {K.~V.}\ \bibnamefont
  {Le}}, \bibinfo {author} {\bibfnamefont {S.}~\bibnamefont {Aya}}, \bibinfo
  {author} {\bibfnamefont {S.}~\bibnamefont {Ogino}}, \bibinfo {author}
  {\bibfnamefont {K.}~\bibnamefont {Okano}}, \bibinfo {author} {\bibfnamefont
  {F.}~\bibnamefont {Araoka}}, \ and\ \bibinfo {author} {\bibfnamefont
  {H.}~\bibnamefont {Takezoe}},\ }\href@noop {} {\bibfield  {journal} {\bibinfo
   {journal} {Molecular Crystals and Liquid Crystals}\ }\textbf {\bibinfo
  {volume} {614}},\ \bibinfo {pages} {124} (\bibinfo {year}
  {2015})}\BibitemShut {NoStop}%
\bibitem [{\citenamefont {Adler}\ \emph
  {et~al.}(2015{\natexlab{a}})\citenamefont {Adler}, \citenamefont {Atherton},
  \citenamefont {Emerson},\ and\ \citenamefont {Mac{L}achlan}}]{Emerson1}%
  \BibitemOpen
  \bibfield  {author} {\bibinfo {author} {\bibfnamefont {J.~H.}\ \bibnamefont
  {Adler}}, \bibinfo {author} {\bibfnamefont {T.~J.}\ \bibnamefont {Atherton}},
  \bibinfo {author} {\bibfnamefont {D.~B.}\ \bibnamefont {Emerson}}, \ and\
  \bibinfo {author} {\bibfnamefont {S.~P.}\ \bibnamefont {Mac{L}achlan}},\
  }\href@noop {} {\bibfield  {journal} {\bibinfo  {journal} {SIAM J. Numer.
  Anal.}\ }\textbf {\bibinfo {volume} {53}},\ \bibinfo {pages} {2226} (\bibinfo
  {year} {2015}{\natexlab{a}})}\BibitemShut {NoStop}%
\bibitem [{\citenamefont {Adler}\ \emph
  {et~al.}(2015{\natexlab{b}})\citenamefont {Adler}, \citenamefont {Atherton},
  \citenamefont {Benson}, \citenamefont {Emerson},\ and\ \citenamefont
  {Mac{L}achlan}}]{Emerson2}%
  \BibitemOpen
  \bibfield  {author} {\bibinfo {author} {\bibfnamefont {J.~H.}\ \bibnamefont
  {Adler}}, \bibinfo {author} {\bibfnamefont {T.~J.}\ \bibnamefont {Atherton}},
  \bibinfo {author} {\bibfnamefont {T.~R.}\ \bibnamefont {Benson}}, \bibinfo
  {author} {\bibfnamefont {D.~B.}\ \bibnamefont {Emerson}}, \ and\ \bibinfo
  {author} {\bibfnamefont {S.~P.}\ \bibnamefont {Mac{L}achlan}},\ }\href@noop
  {} {\bibfield  {journal} {\bibinfo  {journal} {SIAM J. Sci. Comput.}\
  }\textbf {\bibinfo {volume} {37}},\ \bibinfo {pages} {S157} (\bibinfo {year}
  {2015}{\natexlab{b}})}\BibitemShut {NoStop}%
\bibitem [{\citenamefont {Adler}\ \emph {et~al.}(2016)\citenamefont {Adler},
  \citenamefont {Emerson}, \citenamefont {Mac{L}achlan},\ and\ \citenamefont
  {Manteuffel}}]{Emerson3}%
  \BibitemOpen
  \bibfield  {author} {\bibinfo {author} {\bibfnamefont {J.~H.}\ \bibnamefont
  {Adler}}, \bibinfo {author} {\bibfnamefont {D.~B.}\ \bibnamefont {Emerson}},
  \bibinfo {author} {\bibfnamefont {S.~P.}\ \bibnamefont {Mac{L}achlan}}, \
  and\ \bibinfo {author} {\bibfnamefont {T.~A.}\ \bibnamefont {Manteuffel}},\
  }\href@noop {} {\bibfield  {journal} {\bibinfo  {journal} {SIAM J. Sci.
  Comput.}\ }\textbf {\bibinfo {volume} {38}},\ \bibinfo {pages} {B50}
  (\bibinfo {year} {2016})}\BibitemShut {NoStop}%
\bibitem [{\citenamefont {Bangerth}\ \emph {et~al.}(2007)\citenamefont
  {Bangerth}, \citenamefont {Hartmann},\ and\ \citenamefont
  {Kanschat}}]{BangerthHartmannKanschat2007}%
  \BibitemOpen
  \bibfield  {author} {\bibinfo {author} {\bibfnamefont {W.}~\bibnamefont
  {Bangerth}}, \bibinfo {author} {\bibfnamefont {R.}~\bibnamefont {Hartmann}},
  \ and\ \bibinfo {author} {\bibfnamefont {G.}~\bibnamefont {Kanschat}},\
  }\href@noop {} {\bibfield  {journal} {\bibinfo  {journal} {ACM Trans. Math.
  Softw.}\ }\textbf {\bibinfo {volume} {33}},\ \bibinfo {pages} {24/1}
  (\bibinfo {year} {2007})}\BibitemShut {NoStop}%
\bibitem [{\citenamefont {Starke}(2000)}]{Starke1}%
  \BibitemOpen
  \bibfield  {author} {\bibinfo {author} {\bibfnamefont {G.}~\bibnamefont
  {Starke}},\ }\href@noop {} {\bibfield  {journal} {\bibinfo  {journal}
  {Computing}\ }\textbf {\bibinfo {volume} {64}},\ \bibinfo {pages} {323}
  (\bibinfo {year} {2000})}\BibitemShut {NoStop}%
\bibitem [{\citenamefont {Adler}\ \emph {et~al.}(2010)\citenamefont {Adler},
  \citenamefont {Manteuffel}, \citenamefont {McCormick}, \citenamefont {Ruge},\
  and\ \citenamefont {Sanders}}]{adler2010nested}%
  \BibitemOpen
  \bibfield  {author} {\bibinfo {author} {\bibfnamefont {J.~H.}\ \bibnamefont
  {Adler}}, \bibinfo {author} {\bibfnamefont {T.~A.}\ \bibnamefont
  {Manteuffel}}, \bibinfo {author} {\bibfnamefont {S.~F.}\ \bibnamefont
  {McCormick}}, \bibinfo {author} {\bibfnamefont {J.~W.}\ \bibnamefont {Ruge}},
  \ and\ \bibinfo {author} {\bibfnamefont {G.~D.}\ \bibnamefont {Sanders}},\
  }\href@noop {} {\bibfield  {journal} {\bibinfo  {journal} {SIAM J. Sci.
  Comput.}\ }\textbf {\bibinfo {volume} {32}},\ \bibinfo {pages} {1506}
  (\bibinfo {year} {2010})}\BibitemShut {NoStop}%
\bibitem [{\citenamefont {Manteuffel}\ \emph {et~al.}(2006)\citenamefont
  {Manteuffel}, \citenamefont {McCormick}, \citenamefont {Schmidt},\ and\
  \citenamefont {Westphal}}]{manteuffel2006first}%
  \BibitemOpen
  \bibfield  {author} {\bibinfo {author} {\bibfnamefont {T.~A.}\ \bibnamefont
  {Manteuffel}}, \bibinfo {author} {\bibfnamefont {S.~F.}\ \bibnamefont
  {McCormick}}, \bibinfo {author} {\bibfnamefont {J.}~\bibnamefont {Schmidt}},
  \ and\ \bibinfo {author} {\bibfnamefont {C.}~\bibnamefont {Westphal}},\
  }\href@noop {} {\bibfield  {journal} {\bibinfo  {journal} {SIAM J. Numer.
  Anal.}\ }\textbf {\bibinfo {volume} {44}},\ \bibinfo {pages} {2057} (\bibinfo
  {year} {2006})}\BibitemShut {NoStop}%
\bibitem [{\citenamefont {Nocedal}\ and\ \citenamefont
  {Wright}(2006)}]{nocedal2006}%
  \BibitemOpen
  \bibfield  {author} {\bibinfo {author} {\bibfnamefont {J.}~\bibnamefont
  {Nocedal}}\ and\ \bibinfo {author} {\bibfnamefont {S.~J.}\ \bibnamefont
  {Wright}},\ }\href@noop {} {\emph {\bibinfo {title} {Numerical
  Optimization}}}\ (\bibinfo  {publisher} {Springer Verlag},\ \bibinfo {year}
  {2006})\BibitemShut {NoStop}%
\bibitem [{\citenamefont {Logg}\ \emph {et~al.}(2011)\citenamefont {Logg},
  \citenamefont {Mardal}, \citenamefont {Wells} \emph {et~al.}}]{logg2011}%
  \BibitemOpen
  \bibfield  {author} {\bibinfo {author} {\bibfnamefont {A.}~\bibnamefont
  {Logg}}, \bibinfo {author} {\bibfnamefont {K.~A.}\ \bibnamefont {Mardal}},
  \bibinfo {author} {\bibfnamefont {G.~N.}\ \bibnamefont {Wells}},  \emph
  {et~al.},\ }\href@noop {} {\emph {\bibinfo {title} {Automated Solution of
  Differential Equations by the Finite Element Method}}}\ (\bibinfo
  {publisher} {Springer},\ \bibinfo {year} {2011})\BibitemShut {NoStop}%
\bibitem [{\citenamefont {Balay}\ \emph {et~al.}(2015)\citenamefont {Balay},
  \citenamefont {Abhyankar}, \citenamefont {Adams}, \citenamefont {Brown},
  \citenamefont {Brune}, \citenamefont {Buschelman}, \citenamefont {Dalcin},
  \citenamefont {Eijkhout}, \citenamefont {Gropp}, \citenamefont {Kaushik},
  \citenamefont {Knepley}, \citenamefont {McInnes}, \citenamefont {Rupp},
  \citenamefont {Smith},\ and\ \citenamefont {Zhang}}]{balay2015}%
  \BibitemOpen
  \bibfield  {author} {\bibinfo {author} {\bibfnamefont {S.}~\bibnamefont
  {Balay}}, \bibinfo {author} {\bibfnamefont {S.}~\bibnamefont {Abhyankar}},
  \bibinfo {author} {\bibfnamefont {M.~F.}\ \bibnamefont {Adams}}, \bibinfo
  {author} {\bibfnamefont {J.}~\bibnamefont {Brown}}, \bibinfo {author}
  {\bibfnamefont {P.}~\bibnamefont {Brune}}, \bibinfo {author} {\bibfnamefont
  {K.}~\bibnamefont {Buschelman}}, \bibinfo {author} {\bibfnamefont
  {L.}~\bibnamefont {Dalcin}}, \bibinfo {author} {\bibfnamefont
  {V.}~\bibnamefont {Eijkhout}}, \bibinfo {author} {\bibfnamefont {W.~D.}\
  \bibnamefont {Gropp}}, \bibinfo {author} {\bibfnamefont {D.}~\bibnamefont
  {Kaushik}}, \bibinfo {author} {\bibfnamefont {M.~G.}\ \bibnamefont
  {Knepley}}, \bibinfo {author} {\bibfnamefont {L.~C.}\ \bibnamefont
  {McInnes}}, \bibinfo {author} {\bibfnamefont {K.}~\bibnamefont {Rupp}},
  \bibinfo {author} {\bibfnamefont {B.~F.}\ \bibnamefont {Smith}}, \ and\
  \bibinfo {author} {\bibfnamefont {H.}~\bibnamefont {Zhang}},\ }\href
  {http://www.mcs.anl.gov/petsc} {\emph {\bibinfo {title} {{PETS}c Users
  Manual}}},\ \bibinfo {type} {Tech. Rep.}\ \bibinfo {number} {ANL-95/11 -
  Revision 3.6}\ (\bibinfo  {institution} {Argonne National Laboratory},\
  \bibinfo {year} {2015})\BibitemShut {NoStop}%
\bibitem [{\citenamefont {Amestoy}\ \emph {et~al.}(2001)\citenamefont
  {Amestoy}, \citenamefont {Duff}, \citenamefont {Koster},\ and\ \citenamefont
  {L'Excellent}}]{amestoy2001}%
  \BibitemOpen
  \bibfield  {author} {\bibinfo {author} {\bibfnamefont {P.~R.}\ \bibnamefont
  {Amestoy}}, \bibinfo {author} {\bibfnamefont {I.~S.}\ \bibnamefont {Duff}},
  \bibinfo {author} {\bibfnamefont {J.}~\bibnamefont {Koster}}, \ and\ \bibinfo
  {author} {\bibfnamefont {J.-Y.}\ \bibnamefont {L'Excellent}},\ }\href@noop {}
  {\bibfield  {journal} {\bibinfo  {journal} {SIAM Journal on Matrix Analysis
  and Applications}\ }\textbf {\bibinfo {volume} {23}},\ \bibinfo {pages} {15}
  (\bibinfo {year} {2001})}\BibitemShut {NoStop}%
\bibitem [{\citenamefont {Kim}\ \emph {et~al.}(2015{\natexlab{b}})\citenamefont
  {Kim}, \citenamefont {Gim}, \citenamefont {Jung},\ and\ \citenamefont
  {Yoon}}]{YHKim1}%
  \BibitemOpen
  \bibfield  {author} {\bibinfo {author} {\bibfnamefont {Y.~H.}\ \bibnamefont
  {Kim}}, \bibinfo {author} {\bibfnamefont {M.-J.}\ \bibnamefont {Gim}},
  \bibinfo {author} {\bibfnamefont {H.-T.}\ \bibnamefont {Jung}}, \ and\
  \bibinfo {author} {\bibfnamefont {D.~K.}\ \bibnamefont {Yoon}},\ }\href@noop
  {} {\bibfield  {journal} {\bibinfo  {journal} {RSC Adv.}\ }\textbf {\bibinfo
  {volume} {5}},\ \bibinfo {pages} {19279} (\bibinfo {year}
  {2015}{\natexlab{b}})}\BibitemShut {NoStop}%
\bibitem [{\citenamefont {Ackerman}\ \emph
  {et~al.}(2014{\natexlab{b}})\citenamefont {Ackerman}, \citenamefont
  {Trivedi}, \citenamefont {Senyuk}, \citenamefont {van~de Lagemaat},\ and\
  \citenamefont {Smalyukh}}]{Ackerman1}%
  \BibitemOpen
  \bibfield  {author} {\bibinfo {author} {\bibfnamefont {P.~J.}\ \bibnamefont
  {Ackerman}}, \bibinfo {author} {\bibfnamefont {R.~P.}\ \bibnamefont
  {Trivedi}}, \bibinfo {author} {\bibfnamefont {B.}~\bibnamefont {Senyuk}},
  \bibinfo {author} {\bibfnamefont {J.}~\bibnamefont {van~de Lagemaat}}, \ and\
  \bibinfo {author} {\bibfnamefont {I.~I.}\ \bibnamefont {Smalyukh}},\
  }\href@noop {} {\bibfield  {journal} {\bibinfo  {journal} {Phys. Rev. E:
  Stat., Nonlinear, Soft Matter Phys.}\ }\textbf {\bibinfo {volume} {90}}
  (\bibinfo {year} {2014}{\natexlab{b}})}\BibitemShut {NoStop}%
\bibitem [{\citenamefont {Guo}\ \emph {et~al.}(2016{\natexlab{b}})\citenamefont
  {Guo}, \citenamefont {Afghah}, \citenamefont {Xiang}, \citenamefont
  {Lavrentovich}, \citenamefont {Selinger},\ and\ \citenamefont {Wei}}]{Guo1}%
  \BibitemOpen
  \bibfield  {author} {\bibinfo {author} {\bibfnamefont {Y.}~\bibnamefont
  {Guo}}, \bibinfo {author} {\bibfnamefont {S.}~\bibnamefont {Afghah}},
  \bibinfo {author} {\bibfnamefont {J.}~\bibnamefont {Xiang}}, \bibinfo
  {author} {\bibfnamefont {O.~D.}\ \bibnamefont {Lavrentovich}}, \bibinfo
  {author} {\bibfnamefont {R.~L.~B.}\ \bibnamefont {Selinger}}, \ and\ \bibinfo
  {author} {\bibfnamefont {Q.-H.}\ \bibnamefont {Wei}},\ }\href@noop {}
  {\bibfield  {journal} {\bibinfo  {journal} {Soft Matter}\ }\textbf {\bibinfo
  {volume} {12}},\ \bibinfo {pages} {6312} (\bibinfo {year}
  {2016}{\natexlab{b}})}\BibitemShut {NoStop}%
\end{thebibliography}
\end{document}